\documentclass[11pt,fleqn]{article}
\usepackage{epstopdf} 
\usepackage{graphicx}
\usepackage[a4paper,left=2cm,right=2cm]{geometry}
\usepackage{amsmath}
\usepackage{float}
\usepackage[latin1]{inputenc}
\usepackage{amsmath}
\usepackage{amsfonts}
\usepackage{amssymb}
\usepackage{float}
\usepackage[toc,page]{appendix}
\floatstyle{boxed} 
\restylefloat{figure}

\newcommand{\be}{\begin{equation}}
\newcommand{\ee}{\end{equation}}
\newcommand{\bea}{\begin{eqnarray}}
\newcommand{\eea}{\end{eqnarray}}

\begin{document}
\title{Full angular spectrum analysis of tensor current contribution to $A_{cp}(\tau \rightarrow K_{s} \pi \nu_{\tau})$.}
\author{Lobsang Dhargyal. \\\\\ Institute of Mathematical Sciences, Chennai 600113, India and\\\ Harish-Chandra Research Institute, HBNI, Chhatnag Road, Jhusi Allahabad 211 019}

\maketitle
\begin{abstract}

Babar collaboration has reported an intriguing opposite sign in the integrated decay rate asymmetry $A_{cp}(\tau \rightarrow K_{s} \pi \nu_{\tau})$ than that of SM prediction from the known $K^{0}$ - $\bar{K^{0}}$ mixing. Babar's result deviate from the SM prediction by about 2.7$\sigma$. If the result stands with higher precision in the future experiments, the observed sign anomaly in the $A_{cp}(\tau \rightarrow K_{s} \pi \nu_{\tau})$ can most likely come only from a NP. In this work we present a full angular spectrum analysis on the contribution to $A_{cp}(\tau \rightarrow K_{s} \pi \nu_{\tau})$ coming from the tensorial term. Assuming the real part of the NP tensorial coupling is negligible compare to its imaginary part and with $A_{cp}(\tau \rightarrow K_{s} \pi \nu_{\tau})$ and $Br(\tau \rightarrow K_{s} \pi \nu_{\tau})$ as data points to fit the imaginary part of the NP coupling, we have been able to fit the result within 1$\sigma$ of the experimental values.

\end{abstract}

\section{\large Introduction.}

The study of CP violation in tau decays has always been of much interest for beyond the Standard Model studies in the past two decades. In SM, the only source of CP violation is the one phase in the Kobayashi Maskawa (KM) matrix. While the Kobayashi Maskawa ansatz for CP violation within the Standard Model\cite{KM} in the quark sector has been clearly verified by the plethora of data from the B factories, this is unable to account for the observed baryon asymmetry of the Universe. Hence, one needs to look for other sources of CP violation, including searches in the leptonic sector. Apart from the CP phases that may arise in the neutrino mixing matrix, the decays of the tau lepton may allow us to explore nonstandard CP-violating interactions. Various experimental groups have been involved in exploring CP violation in tau decays in the last decade or more. In 2002, the CLEO collaboration\cite{CLEO}, and more recently the Belle Collaboration\cite{Belle2011}, studied the angular distribution of the decay products in $\tau \rightarrow K_{s} \pi \nu_{\tau}$ in search of CP violation; however, neither study revealed any CP asymmetry. The BABAR collaboration\cite{Babar} for the first time reported a sign anomaly in the integrated decay rate asymmetry $A_{cp}(\tau \rightarrow K_{s} \pi \nu_{\tau})$ of
\be
A^{Exp}_{cp} = (-0.36 \pm 0.23 \pm 0.11)\%.\\
\ee
However for $\tau^{\pm} \rightarrow K_{s}^{0} \pi^{\pm} \nu_{\tau} \rightarrow [\pi\pi]^{0}_{K}\pi^{\pm} \nu_{\tau}$, Babar\cite{Babar} has predicted the SM integrated decay-rate asymmetry to be
\be
A^{SM}_{cp} = (0.33 \pm 0.01)\%.\\
\ee
In reference\cite{Grossman}, comparing the rate asymmetries for decays to neutral kaons of the taus with that of D mesons, they have pointed out that since $\tau^{+}(\tau^{-})$ decays initially to a $K^{0}(\bar{K^{0}})$ whereas $D^{+}(D^{-})$ decays initially to $\bar{K^{0}}(K^{0})$, the time-integrated decay-rate CP asymmetry (arising from oscillations of the neutral kaons) of $\tau$ decays must have a sign opposite to that of D decays. The observation of a CP asymmetry in $\tau$ decays to $K_{s}$ having the same sign as that in D decays, and moreover of the same magnitude but opposite in sign to the SM expectation, implies that this asymmetry cannot be accounted for by the CP violation in $K^{0} \bar{K^{0}}$ mixing. Naively one may expect that the simplest way to account for the observed anomaly would be to introduce a direct CP violation via a new CP violating charged scalar exchange. However, it turns out that the charged scalar type of exchange may contribute in the angular distributions, but its mixing with SM term in the integrated decay rate goes to zero. Now the next candidate of NP would be a new CP violating charged vector exchange, but CP violation from vector type NP will be observable only if both vector current and axial vector currents contributes to the same final states\cite{Tsai,Kuhn}. Since in two pseudo scalar meson final states only vector current can contribute due to parity conservation of strong interaction, vector type of NP can contribute in general to CP violation in three or more pseudo scalar meson final states but not in two pseudo scalar meson final states such as $K_{s}\pi$. Now the only possibility left is tensor type of NP.

\section{\large Effective Hamiltonian and decay rates.}

With the assumption that all neutrinos are left handed, we propose the most general effective Hamiltonian containing all possible four fermion interaction operators that can contribute to $\tau \rightarrow K_{s}\pi\nu_{\tau}$  as given by:\\
\be
H_{eff} = \frac{4G_{F}}{\sqrt{2}}V_{us}[(\delta_{l3} + C^{\tau}_{V_{1}})\mathcal{O}^{\tau}_{V_{1}} + C^{\tau}_{V_{2}}\mathcal{O}^{l}_{V_{2}} + C^{\tau}_{S_{1}}\mathcal{O}^{\tau}_{S_{1}} + C^{\tau}_{S_{2}}\mathcal{O}^{\tau}_{S_{2}} + C^{\tau}_{T}\mathcal{O}^{\tau}_{T}] + h.c
\ee
with the operators given by
\be
\mathcal{O}^{\tau}_{V_{1}} = (\bar{s_{L}}\gamma^{\mu}u_{L})(\bar{\nu_{L}}\gamma_{\mu}\tau_{L})\\
\ee
\be
\mathcal{O}^{\tau}_{V_{2}} = (\bar{s_{R}}\gamma^{\mu}u_{R})(\bar{\nu_{L}}\gamma_{\mu}\tau_{L})\\
\ee
\be
\mathcal{O}^{\tau}_{S_{1}} = (\bar{s_{R}}u_{L})(\bar{\nu_{L}}\tau_{R})\\
\ee
\be
\mathcal{O}^{\tau}_{S_{2}} = (\bar{s_{L}}u_{R})(\bar{\nu_{L}}\tau_{R})\\
\ee
\be
\mathcal{O}^{\tau}_{T} = (\bar{s_{L}}\sigma^{\mu\nu}u_{R})(\bar{\nu_{L}}\sigma_{\mu\nu}\tau_{R})\\
\ee
Since we are concern with CP violation in $\tau \rightarrow K_{s}\pi\nu_{\tau}$, we can set the $C^{\tau}_{V_{1}}$ and $C^{\tau}_{V_{2}}$ equal to zero for simplicity as these coefficients will not contribute in CP violation in two meson final states as argued earlier. And as we mentioned earlier and argued in a previous paper of ours \cite{ours} that in the integrated decay rate asymmetry the contribution from the charged scalars goes to zero, so the only terms left is the SM term and the tensor term.\\

\subsection{Decay rate of $\tau \rightarrow K_{s}\pi\nu_{\tau}$ in SM.}

In the SM the $\tau \rightarrow K_{s}\pi\nu_{\tau}$ decay rate can be expressed as:
\be
d\Gamma_{SM}(\tau\rightarrow K\pi\nu) = \frac{1}{2m_{\tau}}\frac{G_F^2}{2}V_{us}^{2}{\cal L_{\mu\nu}H^{\mu\nu}}dPS^{(3)}
\ee
where
\be
{\cal L_{\mu\nu}} =  [{\bar \nu_{\tau}}\gamma_{\mu}(1-\gamma_5)\tau]\,[{\bar \nu_{\tau}}\gamma_{\nu}(1-\gamma_5)\tau]^{\dag}
\ee
and
\be
\cal H^{\mu\nu} = J^{\mu}(J^{\nu})^{\dag}
\ee
where
\be
\mathcal{J^{\mu}} = \langle K(q_{1})\pi(q_2)|V^{\mu}(0)|0 \rangle.
\ee
The hadronic current can be parametrized in terms of the vector and scalar form factors as:
\be
\mathcal{J^{\mu}} = F_V^{K\pi}(Q^2)\big( g^{\mu\nu} - \frac {Q^{\mu}Q{\nu}}{Q^2}\big) (q_1-q_2)_{\nu}\,+ \frac{(m_{K}^{2} - m_{\pi}^{2})}{s}F_{S}^{K\pi}Q^{\nu}\\
\ee
where $Q^{\mu} = (q_{1} + q_{2})^{\mu}$ and in the hadronic rest frame the decay rate can be expressed as:
\be
 \frac{d\Gamma_{SM}(K\pi)}{ds} = \frac{G_F^2V_{us}^{2} m^3_{\tau}}{3 \times 64\pi^3}\frac{1}{s^{\frac{3}{2}}}\left ( 1-\frac{s}{m^2_{\tau}}\right)^2\left (1+\frac{2s}{m^2_{\tau}}\right)\times P(s) \left \{ P(s)^2\,|F_V|^2\,+\,\frac{3(m_k^2-m_{\pi}^2)^2}{4s(1+\frac{2s}{m^2_{\tau}})}|F_S|^2 \right \} 
\label{Gtkpnu}
\ee
where
\be
P(s) = |\vec{q_{1}}| = \frac{1}{2\sqrt{s}}\sqrt{\left[s-(m_k + m_{\pi})^2\right]\left[s-(m_k - m_{\pi})^2\right]}
\ee
is the momentum of the K in the $K\pi$ rest frame and s is the  $K\pi$ invariant mass squared i.e $s = Q^2$. The vector form factor can be parameterized by $K^*(892)$,$K^*(1410)$ and $K^*(1680)$ meson amplitudes given as\cite{Epifanov:2007rf}:
\be
F_V=\frac{1}{1+\beta+\chi}\left[ BW_{K^*(892)}(s) \,+\, \beta BW_{K^*(1410)}(s)\,+\, \chi BW_{K^*(1680)}(s)\right]
\ee
where $\beta$ and $\chi$ are the complex coefficients for the fractions of $K^*(1410)$ and $K^*(1680)$  resonances  respectively and $BW_R(s)$ is a relativistic Breit-Wigner function for R = $K^*(892)$,$K^*(1410)$ and $K^*(1680)$ given as:
\be
BW_R(s)=\frac{M_R^2}{s-M_R^2+i\sqrt{s}\Gamma_R(s)}
\ee
and 
\be
\Gamma_R(s)=\Gamma_{0R}\frac{M_R^2}{s}\left( \frac{P(s)}{P(M_R^2)}\right)^{(2l+1)}
\ee
Here $\Gamma_R(s)$ is the s dependent total width of the resonance and $\Gamma_{0R}(s)$ is the resonance width at its peak and $l=1$ for the vector states and $l=0$ for the s-wave part. Similarly the scalar form factor $F_S$ has $K^*_0(800)$ and $K^*_0(1430)$ contributions and is given as:
\be
F_S=\kappa \frac{s}{M^2_{K^*_0(800)}}BW_{K^*_0(800)}(s)\,+\, \gamma\frac{s}{M^2_{K^*_0(1430)}}BW_{K^*_0(1430)}(s)
\ee
where $\kappa$ and $\gamma$ are the real constants that describe the fractional contributions from  $K^*_0(800)$ and $K^*_0(143
0)$ respectively. As reported by Belle\cite{Epifanov:2007rf}, $K^*_{(892)}$ alone is not enough to describe the $K_s \pi$ mass spectrum. It is best explained for $K^*(892)+ K^*(1410) + K^*(800)$  and   $K^*(892)+ K^*(1430) + K^*(800)$. We will use $K^*(892)+ K^*(1410) + K^*(800)$ in this analysis which best fits the Belle mass spectrum.

\subsection{Tensorial term.}

We now include the contribution from the tensorial operator as it has been already pointed out earlier that scalar and the vectorial operators would not contribute to the integrated decay rate asymmetry and CPV. The key requirement in the relevant context of explaining the observed CPV in integrated $\tau \rightarrow K_{s}\pi\nu_{\tau}$ decay rate by the tensorial operator is that its coefficient $C^{l}_{T}$ from Eqs.(8) should be complex so that interference of the SM with this tensor amplitude gives the required CP phase. We have from Eqs.(3) the effective Hamiltonian given as\\
\be
\mathcal{H}_{eff}^{T} = \frac{4G_{F}}{\sqrt{2}}V_{us}C^{\tau}_{T}(\bar{s_{L}}\sigma^{\mu\nu}u_{R})(\bar{\nu_{L}}\sigma_{\mu\nu}\tau_{R})\\
\ee
where $\sigma^{\mu\nu} = \frac{i}{2}(\gamma^{\mu}\gamma^{\nu} - \gamma^{\nu}\gamma^{\mu})$ and the hadronic current can be expressed as\\
\be
\langle K(q_{1})\pi(q_2)|\bar{s}\sigma^{\mu\nu}u|0 \rangle = i\frac{2F_{T}}{m_{K} + m_{\pi}}(q^{\mu}_{1}q^{\nu}_{2} - q^{\mu}_{2}q^{\nu}_{1}).\\
\ee
where $F_{T}$ is the tensorial form factor and only tensor term can contribute due to parity conservation of strong interaction and pseudo-tensor term will not contribute. In a previous collaboration involving the author\cite{ours}, we have argued that tensor type of NP may be able to explain the observed sign anomaly however in that work we have assumed that the tensor form factors are constants, but it turns out that is not the case in general and so in this work we have been able to express the tensor form factors in terms of scalar and vector form factors using Dirac equations of motion.\\
We have from the equations of the motion:
\be
\partial_{\nu}(\bar{u}_{s}\sigma^{\mu\nu}v_{\bar{u}}) = (m_{s} + m_{u})\bar{u}_{s}\gamma^{\mu}v_{\bar{u}} + (i\partial_{\mu}\bar{u}_{s})v_{\bar{u}} - \bar{u}_{s}(i\partial_{\mu} v_{\bar{u}})
\label{eq:eqmotion}
\ee
which gives
\be
iQ_{\nu}\langle K(q_{1})\pi(q_2)|\bar{u}_{s}\sigma^{\mu\nu}v_{\bar{u}}|0 \rangle = -[-(m_{s} + m_{u})\langle K(q_{1})\pi(q_2)|\bar{u}_{s}\gamma^{\mu}v_{\bar{u}}|0 \rangle + \langle K(q_{1})\pi(q_2)|\bar{u}_{s}v_{\bar{u}}|0 \rangle M(q_{1}-q_{2})^{\mu}]
\label{eq:qrk-meson}
\ee
Where we define $\langle K(q_{1})\pi(q_2)|\bar{s}u|0 \rangle =  F_{0}$ with M an adjustable parameter and now contracting Eqs(13) from section 2.1 with $Q_{\mu}$ we get $F_{0} = +\frac{(m_{K}^{2} - m_{\pi}^{2})}{(m_{s}-m_{u})}F_{S}$ where $Q_{\nu} = (q_{1} + q_{2})_{\nu}$. Our justification in going from Eqs(\ref{eq:eqmotion}) to Eqs(\ref{eq:qrk-meson}) is that since strong force is mass independent, the corrections to replacing the quark four momentum with respective meson four momentum would be same to both s and u quarks and so it would be a common factor (M) and all other factors absorbed into the form factors.  Now using Eqs(13,21) and the $F_{0}$ given above, after few algebraic manipulations we can express the tensor form factor $F_{T}$ in terms of scalar form factor $F_{S}$ and vector form factor $F_{V}$ by comparing the coefficients of $Q^{\mu}$ and $(q_{1}-q_{2})^{\mu}$ from LHS and RHS of Eqs(23), details can be found in the appendrix, which gives

\be
F^{a}_{T} = \frac{(m_{s} + m_{u})(m_{K} + m_{\pi})}{s}[ -F_{V} + F_{S} ]
\ee
and\\
\be
F^{b}_{T} = -\frac{(m_{s} + m_{u})(m_{K} + m_{\pi})}{s}[ F_{V} - \frac{(m_{K}^{2} - m_{\pi}^{2})}{(m_{s}^{2} - m_{u}^{2})}M F_{S} ]\\
\ee
We fix M such that $F^{b}_{T} = F^{a}_{T}$, from the forms of $F^{a}_{T}$ and $F^{b}_{T}$, if we require $M = \frac{(m_{s}^{2} - m_{u}^{2})}{(m_{K}^{2} - m_{\pi}^{2})}$, then clearly $F^{b}_{T} = F^{a}_{T}$. This value of M seems to be a reasonable measure of Quark-Hadron duality violation in these kind of reactions, where in the Quark-Hadron duality limit, M $\rightarrow$ 1.\footnote{the reason why we neglected a correction factor, similar to M, when replacing total quark momentum ($q_{1} + q_{2}$) with total hadron momentum ($Q$) in the LHS of Eqs.(\ref{eq:qrk-meson}) is because it goes through $K^{*}$ resonances,a QCD bound state, where most of the energy momentum of the resonance is expected to be carried by the quarks (as only soft gloun exchange between u and s quarks are expected to dominate due to larger $\alpha_{s}(Q^{2})$ at low $Q^{2}$)...} See Figure \ref{Fig1:fig1} for the plot of $|F^{a}_{T}|$ as a function of hadronic invariant mass squared.

\subsection{Including the contribution from the tensor term to the $\tau \rightarrow K_{s}\pi\nu_{\tau}$ decay rate.}

When tensorial term is included the total decay rate is given by\\
\be
d\Gamma = (\frac{d\Gamma_{SM}}{ds} + \frac{d\Gamma_{MIX}}{ds} + \frac{d\Gamma_{T}}{ds})ds
\ee
where the $\frac{d\Gamma_{SM}}{ds}$ is given in the Eqs.(14) and the full angular dependence of the other two terms can be expressed as:\\
\be
\begin{split}
\frac{d\Gamma_{MIX}}{ds\frac{d\cos{\beta}}{2}\frac{d\alpha}{2\pi}} = -\frac{G_F^2V_{us}^{2} m^2_{\tau}}{\pi^3 (m_{k} + m_{\pi}) 2s^{\frac{1}{2}}}( 1-\frac{s}{m^2_{\tau}})^2 P^2 \{ -P\times Re(F_{V}^{\dagger}F_{T}C_{T}) + Re(F_{S}^{\dagger}F_{T}C_{T})\\
\times [ \frac{m_{k}^{2} - m_{\pi}^{2}}{2\sqrt{s}} ]\times (\sin{\beta}\cos{\alpha}\sin{\psi} + \cos{\beta}\cos{\psi}) \}
\label{eq:mix}
\end{split}
\ee
and\\
\be
\begin{split}
\frac{d\Gamma_{T}}{ds\frac{d\cos{\beta}}{2}\frac{d\alpha}{2\pi}} = \frac{G_F^2V_{us}^{2} m_{\tau}^{3}|F_{T}|^{2}}{(m_{k} + m_{\pi})^{2}\pi^3 2s^{\frac{1}{2}}}( 1-\frac{s}{m^2_{\tau}})^2 P^2 \{ \frac{P}{2} + \frac{3}{2}(s - m_{k}^{2} - m_{\pi}^{2})\frac{(m_{k}^{2} - m_{\pi}^{2})}{s^{3/2}}\\
\times (\sin{\beta}\cos{\alpha}\sin{\psi} + \cos{\beta}\cos{\psi}) - (1 - \frac{s}{m_{\tau}^{2}})\frac{P}{2}\\
\times (\sin{\beta}\cos{\alpha}\sin{\psi} + \cos{\beta}\cos{\psi})^{2} \}
\label{eq:ten}
\end{split}
\ee
Where the P is same as in Eqs(15)and the angles $\alpha$, $\beta$ are same as defined in Figure 1 of reference \cite{Kuhn2} and $\psi$ is defined as the angle between direction of flight of the lab frame and the direction of flight of $\tau$ as seen from the hadronic rest frame. We now integrate over the $\cos{\beta}$ from -1 to +1 and $\alpha$ from 0 to 2$\pi$, and require that $Re(C^{\tau}_{T}) << Im(C^{\tau}_{T})$ to avoid too large NP contribution to $Br(\tau \rightarrow K_{s}\pi\nu_{\tau})$ which has been measured with much more accurately then $A_{cp}(\tau \rightarrow K_{s}\pi\nu_{\tau})$, so then we can approximately take $Re(C^{\tau}_{T}) \approx 0$ and we are left with only one parameter $Im(C^{\tau}_{T})$ to fit. We can now use the $A_{cp}(\tau \rightarrow K_{s}\pi\nu_{\tau})$ and $Br(\tau \rightarrow K_{s}\pi\nu_{\tau})$ as data points to fit the $Im(C^{\tau}_{T})$ parameter. In a previous collaboration involving the author\cite{ours}, we have shown in the Eqs(44) of that reference that the CPV coming from the $K-\bar{K}$ mixing and the direct CPV in $A_{cp}(\tau \rightarrow K_{s}\pi\nu_{\tau})$ can be seperated as\\
\be
A_{cp}(\tau \rightarrow K_{s}\pi\nu_{\tau}) = \frac{A_{cp}^{K} + A_{cp}^{\tau}}{1 + A_{cp}^{K}A_{cp}^{\tau}}
\ee
and also we have
\be
Br(\tau \rightarrow K_{s}\pi\nu_{\tau}) = \frac{(\Gamma^{\tau^{+}} + \Gamma^{\tau^{-}})}{2}\tau_{\tau}
\ee
where $A_{cp}^{K}$ is the CPV coming from the $K-\bar{K}$ mixing and $A_{cp}^{\tau}$ is the direct CP violation coming from NP particle mediated CPV at lepton and/or quark vertices and $\tau_{\tau}$ is the $\tau$ life time. Since both $A_{cp}^{K}$ and $A_{cp}^{\tau}$ are expected to be small, we can savely ignore terms involving the product of the two. And also since $Re(C^{\tau}_{T}) \approx 0$ and the sign of the complex part is opposite in $\Gamma^{\tau^{+}}$ relative to the $\Gamma^{\tau^{-}}$, the branching fraction receives no contribution from the SM and Tensorial mixing part.

\section{Results.}

\begin{figure}[!h]
\begin{minipage}[t]{0.48\textwidth}
\hspace{0.4cm}
\includegraphics[width=1.8\linewidth, height=9.2cm]{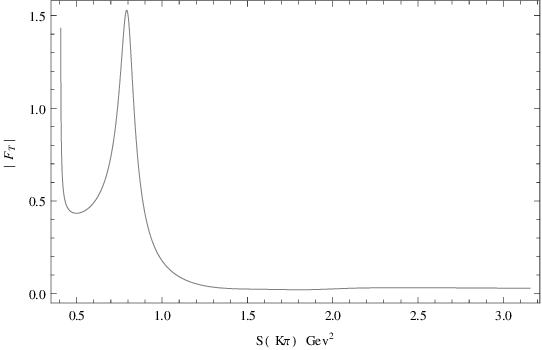}
\end{minipage}
\caption{This figure shows the plots of $|F^{a}_{T}|$ as a function of $S(K\pi)$ where $S(K\pi)$ is the hadronic invariant mass squared.}
\label{Fig1:fig1}
\end{figure}

With taking the approximation of $A^{k}_{pc}A^{\tau}_{cp} \approx 0$ we can express Eqs(29,30) as:
\be
A_{cp}(\tau \rightarrow K_{s}\pi\nu_{\tau}) = A_{cp}^{K} + A_{cp}^{\tau}
\ee
and
\be
Br(\tau \rightarrow K_{s}\pi\nu_{\tau}) = \frac{(\Gamma^{\tau^{+}} + \Gamma^{\tau^{-}})}{2}\tau_{\tau} = (\Gamma_{SM} + \Gamma_{T})\tau_{\tau} 
\ee
where $A^{k}_{cp}$ is the known SM CPV from the $K-\bar{K}^{0}$ mixing, $\Gamma_{SM}$ is the SM decay rate corresponding to fitted form factors from Belle\cite{Epifanov:2007rf}, $\Gamma_{T}$ is the tensorial decay rate we gets from integration of Eqs(28) and $\tau_{\tau}$ is the life time of $\tau$ lepton. From Eqs(31,32) and using $F^{a}_{T}$ from Eqs(24) the best fitted value of the complex parameter $Im(C^{\tau}_{T})$ to the two data points gives at $\chi^{2} \approx 4.5$ :
\be
Im(C^{\tau}_{T}) =  -0.071,
\label{eq:Ct}
\ee
which gives
\be
Br(\tau \rightarrow K_{0}\pi\nu_{\tau})^{(Th)} = 2Br(\tau \rightarrow K_{s}\pi\nu_{\tau})^{(Th)} = (0.756 \pm 0.085)\%\\
\ee
and
\be
A^{\tau(Th)}_{cp} = (-0.703 \pm 0.54)\%\\
\ee
whereas the experimental values of these observables are given as 
\be
A^{(Exp-SM)}_{cp} = A^{\tau(Exp)}_{cp} - (A^{k}_{cp})^{SM} = (-0.69 \pm 0.26)\%,\\
\ee
and
\be
Br(\tau \rightarrow K_{0}\pi\nu_{\tau})^{(Exp)} = 2Br(\tau \rightarrow K_{s}\pi\nu_{\tau})^{(Exp)} = (0.84 \pm 0.04)\%.\\
\ee
Comparing Eqs(35,36) and Eqs(34,37) we see that the theoretical predicted values fit with the experimental values within 1$\sigma$. 
In Figure~\ref{Fig1:fig1} we have shown the plots of $|F^{a}_{T}|$ as a function of $S(K\pi)$ where $S(K\pi)$ is the hadronic invariant mass squared.

\section{Future Prospects.}

\begin{figure}[h!]
\begin{minipage}[t]{0.48\textwidth}
\hspace{0.4cm}
\includegraphics[width=1.8\linewidth, height=7.2cm]{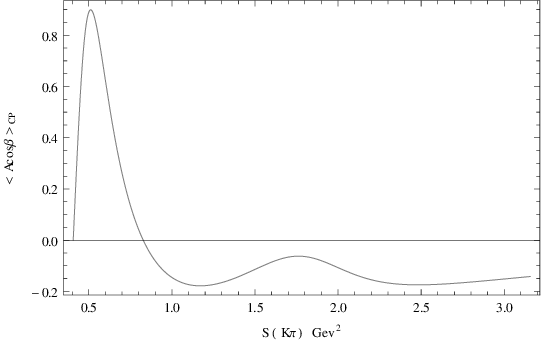}
\end{minipage}
\caption{This figure shows the plots of $\langle A \cos{\beta} \rangle_{CP}(S)$ as a function of $S(K\pi)$ where $S(K\pi)$ is the hadronic invariant mass squared and $Im(C_{T})$ is taken from Eqs(\ref{eq:Ct}).}
\label{Fig1:fig2}
\end{figure}

In what follows we will assume the the direction of the $\tau$ has been measured and so we can set $\psi \rightarrow 0$ in Eqs(\ref{eq:mix}) and Eqs(\ref{eq:ten}). Then since all the terms which depends on $\alpha$ goes to zero, we can integrate out in $\alpha$ also. Now then the angular dependence of the mixing term is given as:
\be
\begin{split}
\frac{d\Gamma_{MIX}}{ds\frac{d\cos{\beta}}{2}\frac{d\alpha}{2\pi}} = -\frac{G_F^2V_{us}^{2} m^2_{\tau}}{\pi^3 (m_{k} + m_{\pi}) 2s^{\frac{1}{2}}}( 1-\frac{s}{m^2_{\tau}})^2 P^2 \{ -P\times Re(F_{V}^{\dagger}F_{T}C_{T}) + Re(F_{S}^{\dagger}F_{T}C_{T})\\
\times [ \frac{m_{k}^{2} - m_{\pi}^{2}}{2\sqrt{s}} ]\times (\cos{\beta}) \}.
\label{eq:mix2}
\end{split}
\ee
It is clear from Eqs(\ref{eq:mix}) and Eqs(\ref{eq:mix2}) that the mixing of the vector form factor($F_{V}$) with the tensor form factor($F_{T}$) has no dependence on any of the angles, all angular dependence cancels, and so CP violation coming from the interference of the vector part of SM current and the New Tensor current will show up in angular integrated decay rate as we found in previous section. And also from Eqs(\ref{eq:mix}) and Eqs(\ref{eq:mix2}) we notice that the CP violation coming from the interference of the scalar part of the SM current and the New Tensor current will not contribute to angular integrated CP violation and decay rate, but it can contribute in the angular distribution spectrum. One simplest way to extract the angular dependence, especially in the case of linear dependence ones like in Eqs(\ref{eq:mix2}), is by weighted integrals. We will use $\cos{\beta}$ as weight multiplying the differential rate and then integrate out in $d\cos{\beta}/2$ given as:
\be
\begin{split}
\frac{d}{ds}[ \langle \frac{\Gamma_{MIX} \times \cos{\beta}}{d\cos{\beta}/2} \rangle - \langle \frac{\bar{\Gamma}_{MIX} \times \cos{\beta}}{d\cos{\beta}/2} \rangle ] = -\frac{G_{F}^{2}|V_{us}|^{2}m_{\tau}^{2}(m_{k} - m_{\pi})}{6\pi^{3}}Im[C_{T}](1 - \frac{s}{m_{\tau}^{2}})^{2}\frac{P^{2}}{s}Im[F_{s}^{\dagger}F_{T}],
\label{eq:wght}
\end{split}
\ee
then by normalizing Eqs(\ref{eq:wght}) by $\frac{1}{2}( \frac{d\Gamma}{ds} + \frac{d\bar{\Gamma}}{ds} )$ we have
\be
\begin{split}
\langle A \cos{\beta} \rangle_{CP}(S) = \frac{ \frac{d}{ds}[ \langle \frac{\Gamma_{MIX} \times \cos{\beta}}{d\cos{\beta}/2} \rangle - \langle \frac{\bar{\Gamma}_{MIX} \times \cos{\beta}}{d\cos{\beta}/2} \rangle ] }{ \frac{1}{2}( \frac{d\Gamma}{ds} + \frac{d\bar{\Gamma}}{ds} ) }.
\label{eq:AcpB}
\end{split}
\ee
In Figure \ref{Fig1:fig2} we have shown the plot of $\langle A_{CP} \cos{\beta} \rangle$ as a function of $S(K_{s}\pi)$ using $Im(C_{T})$ from Eqs(\ref{eq:Ct})

\newpage
Now by integrating in the $S(K\pi)$ in the range ($(m_{k} + m_{\pi})^{2}$,\ $m_{\tau}^{2}$) we have:
\be
\begin{split}
\int^{m_{\tau}^{2}}_{(m_{k} + m_{\pi})^{2}} ( \langle A \cos{\beta} \rangle_{CP}(S) )dS = -0.127.
\label{eq:AcpInt}
\end{split}
\ee
If the observed anomaly in the $A_{CP}$ is due to a new tensor interaction, then from the above equation we can expect quite large CP violation to be observed in the angular weighted CP asymmetry in the $\tau \rightarrow K_{s}\pi\nu_{\tau}$ decay mode in future experimental searches. As we can see from the Figure \ref{Fig1:fig2}, CP violation from new tensor interaction will show up most dramatically in the low hadronic invariant mass square($S(K_{S}\pi)$) region.

\section{Conclusions.}

Babar collaboration has reported an intriguing opposite sign in the integrated decay rate asymmetry $A_{cp}(\tau \rightarrow K_{s} \pi \nu_{\tau})$ than that of SM prediction from the known $K^{0}$ - $\bar{K^{0}}$ mixing. Babar's result deviate from the SM prediction by about 2.7$\sigma$. In this work we have presented an improved analysis of our previous work on the contributions coming from tensorial current to this observable. Assuming the real part of the NP coupling is negligible compare to its imaginary part, the best fitted value of the parameter $Im(C^{\tau}_{T})$ to the two data points $A_{cp}(\tau \rightarrow K_{s} \pi \nu_{\tau})$ and $B_{r}(\tau \rightarrow K_{0} \pi \nu_{\tau})$ is given by $Im(C^{\tau}_{T}) = -0.071$ which gives $A^{Th}_{cp} = (-0.703 \pm 0.51)\times10^{-2}$ compare to the experimental minus SM value of $A^{(Exp-SM)}_{cp} = (A^{Exp}_{cp} - A_{cp}^{SM}) = (-0.69 \pm 0.26)\times10^{-2}$. And similarly we have $Br(\tau \rightarrow K_{0}\pi\nu_{\tau})^{(Th)} = (0.756 \pm 0.084)\times10^{-2}$ comapre to the $Br(\tau \rightarrow K_{0}\pi\nu_{\tau})^{(Exp)} = (0.84 \pm 0.04)\times10^{-2}$. As we can see the theoretical predictions fit with the experimental results within 1$\sigma$ for both observables. If the observed anomaly in the $A_{CP}$ is due to a new tensor interaction, then according to Eqs(\ref{eq:AcpInt}), we can expect quite large CP violation to be observed in the angular weighted CP asymmetry in the $\tau \rightarrow K_{s}\pi\nu_{\tau}$ decay mode in future experimental searches.\\

{\large Acknowledgments: \large} This work is supported and funded by the Department of Atomic Energy of the Government of India and by the Government of Tamil Nadu and U.P.

\newpage
\begin{appendices}

\section{Expressing $F_{T}$ in Terms of $F_{V}$ And $F_{S}$ Using Equations of Motion.}
We have from the equations of the motion:
\be
\partial_{\nu}(\bar{u}_{s}\sigma^{\mu\nu}v_{\bar{u}}) = (m_{s} + m_{u})\bar{u}_{s}\gamma^{\mu}v_{\bar{u}} + (i\partial_{\mu}\bar{u}_{s})v_{\bar{u}} - \bar{u}_{s}(i\partial_{\mu} v_{\bar{u}})
\label{eq:epdx1}
\ee
which gives
\be
iQ_{\nu}\langle K(q_{1})\pi(q_2)|\bar{u}_{s}\sigma^{\mu\nu}v_{\bar{u}}|0 \rangle = -[-(m_{s} + m_{u})\langle K(q_{1})\pi(q_2)|\bar{u}_{s}\gamma^{\mu}v_{\bar{u}}|0 \rangle + \langle K(q_{1})\pi(q_2)|\bar{u}_{s}v_{\bar{u}}|0 \rangle M (q_{1}-q_{2})^{\mu}]
\label{eq:epdx2}
\ee
Our justification in going from Eqs(\ref{eq:epdx1}) to Eqs(\ref{eq:epdx2}) is that since strong force is mass independent, the corrections in replacing the quark four momentum with respective meson four momentum would be same to both s and u quarks and so it would be a common factor (M) and all other factors absorbed into the form factors. So then we have
\be
-Q_{\nu}\frac{2F_{T}}{m_{K} + m_{\pi}}(q^{\mu}_{1}q^{\nu}_{2} - q^{\mu}_{2}q^{\nu}_{1}) = -[-(m_{s} + m_{u}) \langle K(q_{1})\pi(q_{2})| \bar{s}\gamma^{\mu}u |0 \rangle + M F_{0}(q_{1} - q_{2})^{\mu}]
\ee
with $\langle K(q_{1})\pi(q_2)|\bar{u}_{s}v_{\bar{u}}|0 \rangle = F_{0}$ where M is an adjustable parameter and $Q_{\nu} = (q_{1} + q_{2})_{\nu}$.\footnote{for $\langle K(q_{1})\pi(q_2)|\bar{u}_{u}v_{\bar{s}}|0 \rangle$ we have $F_{0} = -\frac{(m_{K}^{2} - m_{\pi}^{2})}{(m_{s}-m_{u})}F_{S}$ but that minus sign is compensated by a negative sign in second term in RHS of Eqs.(\ref{eq:eqmotion})(charge conjugated one)} Using $\langle K(q_{1})\pi(q_{2})| \bar{u}_{s}\gamma^{\mu}v_{\bar{u}} |0 \rangle = F_V^{K\pi}(Q^2)\big( g^{\mu\nu} - \frac {Q^{\mu}Q{\nu}}{Q^2}\big) (q_1-q_2)_{\nu}\,+ \frac{(m_{K}^{2} - m_{\pi}^{2})}{s}F_{S}^{K\pi}Q^{\nu}$\footnote{this is intuitively understood as $\langle K(q_{1})\pi(q_{2})| |K^{*}(892);K^{*}(1430)_{0} \rangle \langle K^{*}(892);K^{*}(1430)_{0}| a^{\dagger}_{s}b^{\dagger}_{\bar{u}} \bar{u}_{s}\gamma^{\mu}v_{\bar{u}} |0 \rangle $, where the negative sign from the antiparticle wave function under parity transformation is canceled by the negative sign under parity transformation for the antiparticle creation operator, hence the current as a whole behave like a vector under parity.} and contracting it with $Q_{\nu}$ gives $F_{0} = +\frac{(m_{K}^{2} - m_{\pi}^{2})}{(m_{s}-m_{u})}F_{S}$, and using the identity $Q\cdot q_{2} q^{\mu}_{1} - Q\cdot q_{1} q^{\mu}_{2} = \frac{(Q\cdot q_{1} + Q\cdot q_{2})(q_{1} - q_{2})^{\mu} - (Q\cdot q_{1} - Q\cdot q_{2})Q^{\mu}}{2}$ we have
\be
\begin{split}
\frac{F_{T}}{m_{K} + m_{\pi}} [ (Q\cdot q_{1} + Q\cdot q_{2})(q_{1} - q_{2})^{\mu} - (Q\cdot q_{1} - Q\cdot q_{2})Q^{\mu} ] =\\
[-(m_{s} + m_{u})F_{V} + \frac{(m_{K}^{2} - m_{\pi}^{2})}{(m_{s}-m_{u})}M F_{S}](q_{1} - q_{2})^{\mu}\\
- [ -(m_{s} + m_{u})(m_{K}^{2}- m_{\pi}^{2})/Q^{2}F_{V} + (m_{s} + m_{u})(m_{K}^{2} - m_{\pi}^{2})/Q^{2}F_{S} ]Q^{\mu}
\end{split}
\label{eq:epdx4}
\ee
where $Q\cdot q_{1} = \frac{Q^{2} + m^{2}_{K} - m^{2}_{\pi}}{2}$ and $Q\cdot q_{2} = \frac{Q^{2} + m^{2}_{\pi} - m^{2}_{K}}{2}$; then comparing the coeffecients of $(q_{1} + q_{2})^{\mu}$ and $(q_{1} - q_{2})^{\mu}$ from the LHS and RHS of Eqs(41) we have,
\be
F^{a}_{T} = \frac{(m_{s} + m_{u})(m_{K} + m_{\pi})}{s}[ -F_{V} + F_{S} ]
\ee
and\\
\be
F^{b}_{T} = -\frac{(m_{s} + m_{u})(m_{K} + m_{\pi})}{s}[ F_{V} - \frac{(m_{K}^{2} - m_{\pi}^{2})}{(m_{s}^{2} - m_{u}^{2})}M F_{S} ]\\
\ee
Now to fix M we contract Eqs.(\ref{eq:epdx4}) by $Q_{\mu}$, then the LHS gives zero and the RHS gives $M = \frac{(m_{s}^{2} - m_{u}^{2})}{(m_{K}^{2} - m_{\pi}^{2})}$, which when put in $F^{b}_{T}$, shows that $F^{a}_{T} = F^{b}_{T}$. Contracting Eqs.(\ref{eq:epdx4}) with $(q_{2} - q_{1})_{\mu}$ will give,using $M = \frac{(m_{s}^{2} - m_{u}^{2})}{(m_{K}^{2} - m_{\pi}^{2})}$, $F_{T}$ same as $F^{a}_{T}$ in Eqs.(42).

\end{appendices}

\end{document}